\documentclass[12pt]{article}

\usepackage{lmodern}
\usepackage[T2A,T1]{fontenc}
\usepackage[utf8]{inputenc}
\usepackage{geometry}
\usepackage{bm}
\usepackage{mathtools}
\usepackage{amsmath}
\usepackage{amssymb}
\usepackage{graphicx}
\usepackage{textcomp}
\usepackage[numbers,sort&compress]{natbib}
\usepackage{hyperref}
\usepackage{color}
\hypersetup{colorlinks=true,citecolor=blue,linkcolor=black,urlcolor=black}
\numberwithin{equation}{section}

\def\Ord{{\mathcal O}}
\def\Operator{{\mathcal O}}
\def\Operator{{\mathbb O}}

\def\id{\protect{{1 \kern-.28em {\rm l}}}}
\def\pol{\varepsilon}

\newcommand{\eqn}[1]{Eq.~\eqref{#1}}

\def\C2{{C_{ES^2}}}
\def\CB3{{C_{BS^3}}}

\def\vq{\bm q}
\def\vb{\bm b}
\def\vp{\bm p}
\def\vpp{{\vp^2}}
\def\vS1{\bm S_1}

\def\pd{\partial}
\def\nn{\nonumber}

\def\cO{\mathcal{O}}
\def\cE{\mathcal{E}}

\def\S{{{\mathbb S}}}
\def\pS{{\hat {\bm{S}}}}
\def\Vmom{\iffalse\widehat{V}\fi V}
\def\FTLq{\bm{L}_{q}}
\def\FTLQ{\bm{L}_{\hat{q}}}

\newcommand{\be}{\begin{equation}}
	\newcommand{\ee}{\end{equation}}

\allowdisplaybreaks
\begin{document}
	\interfootnotelinepenalty=10000
	\baselineskip=18pt
	\hfill
	
	\thispagestyle{empty}

% UCLA/TEP/2020/102
	\vspace{1.2cm}

	\begin{center}
		{ \bf \Large 
Quadratic-in-Spin Hamiltonian at $\mathcal{O}(G^2)$ \\ from Scattering Amplitudes
		}

		\bigskip\vspace{1.cm}{
			{\large 
		Dimitrios Kosmopoulos and Andres Luna
		}
		} \\[7mm]
		{\it  
			Mani L. Bhaumik Institute for Theoretical Physics, \\[-1mm]
			  Department of Physics and Astronomy, UCLA, Los Angeles, CA 90095 
		}
                  \\
	\end{center}
	\bigskip
	\bigskip

	\begin{abstract} \small

We obtain the quadratic-in-spin terms of the conservative
Hamiltonian describing the interactions of a binary of spinning bodies in General Relativity
through $\mathcal{O}(G^2)$ and to all orders in velocity.
Our calculation extends a recently-introduced framework based on scattering amplitudes and effective field
theory to consider non-minimal coupling of the spinning objects to gravity.
At the order that we consider, we establish the validity of the formula  proposed in \cite{Bern:2020buy} that relates  the impulse and spin kick in a scattering event to the eikonal phase.

\end{abstract}

	\setcounter{footnote}{0}
	
\renewcommand{\baselinestretch}{1}	
	\newpage
	\setcounter{tocdepth}{2}
	\tableofcontents
	
	\newpage

%%%%%%%%%%%%%%%%%%%%%%%%%%%%%%%%%%%%%%%%%%%%%%%%%%%%%%%%%%%%%%%

\section{Introduction}
\label{sec:intro}

The detection of gravitational waves by the LIGO and Virgo collaborations~\cite{Abbott:2016blz,TheLIGOScientific:2017qsa} promises intriguing new discoveries. 
The main sources of gravitational waves are binary systems of compact astrophysical objects. Therefore, the great experimental advances also press for the development of high-precision theoretical tools for the modeling of the evolution of such systems.
In the present paper we consider the inspiral phase of the evolution of the binary. A well-developed theoretical tool to study this phase is the post-Newtonian (PN) approximation. 
%
%In the inspiral phase which concerns the present paper,  one such tool is the post-Newtonian (PN) approximation. 
This approach consists of an expansion in small velocities and weak gravitaional field.  Several methods based on General Relativity (GR)~\cite{Blanchet:2013haa,Schafer:2018kuf} as well as effective field theory (EFT)~\cite{Goldberger:2004jt} have been developed in this direction.
We instead choose to use the post-Minkowskian (PM) approximation,  an expansion in Newton's constant $G$ which yields the exact velocity dependence.  The PM approximation has a long history in GR ~\cite{Westpfahl:1985tsl} but has gained prominence recently (see e.g.~\cite{Damour:2016gwp,Damour:2017zjx,Damour:2019lcq}) due in part to the successful adaptation of modern scattering-amplitudes techniques.

The application of quantum-field-theory (QFT) methods to the study of the two-body problem dates back to the 1970's~\cite{Iwasaki:1971vb}. However,  it was recently that Ref. ~\cite{Neill:2013wsa} proposed the application of the well-established scattering-amplitudes toolkit to the  derivation of gravitational potentials (see Refs. ~\cite{Dixon:1996wi,Elvang:2013cua,Bern:2019prr} for reviews on the modern amplitudes program).
Along these lines,  Ref. ~\cite{Cheung:2018wkq} developed an EFT of non-relativistic scalar fields which allowed the construction of the 2PM\footnote{The $n$PM order corresponds to $\cO(G^n)$.} canonical Hamiltonian from a one-loop scattering amplitude. This Hamiltonian was equivalent to the one 
of Ref.~\cite{Westpfahl:1985tsl}.  
Refs.~\cite{Bern:2019nnu,Bern:2019crd,Cheung:2020gyp} later implemented this approach to obtain novel results at 3PM order.  Ref.~\cite{Antonelli:2019ytb} followed up shortly after to compare these results against numerical relativity in terms of the energetics of the binary. Very recently, Ref. ~\cite{Bern:2021dqo} obtained the conservative binary potential at 4PM order.

Besides making use of a non-relativistic EFT, various approaches have been developed to extract the dynamics of compact non-spinning objects from scattering data.  Refs.~\cite{Kosower:2018adc,delaCruz:2020bbn} established a formalism to obtain physical observables from unitarity cuts.  Refs.~\cite{Cristofoli:2019neg,Bjerrum-Bohr:2019kec} made use of the Lippman-Schwinger equation.  Refs.~\cite{Kalin:2019rwq,Kalin:2019inp} developed a boundary-to-bound (B2B) dictionary, and Refs.~\cite{Kalin:2020mvi,Kalin:2020fhe} implemented a worldline PM EFT.   Ref. ~\cite{Bern:2021dqo} discovered an amplitude-action relation that allows the calculation of physical observables directly from the scattering amplitude.

The techniques mentioned above have been extended in multiple directions in recent years. Indeed,  Refs.~\cite{Caron-Huot:2018ape,Bern:2020gjj,Parra-Martinez:2020dzs} applied similar methods to supergravity.  Ref.~\cite{Loebbert:2020aos} studied three-body dynamics, while Refs.~\cite{DiVecchia:2020ymx,DiVecchia:2021ndb,Herrmann:2021lqe,Damour:2020tta} incorporated the radiation emitted by the binary into their analysis.  Refs.~\cite{Haddad:2020que,Aoude:2020ygw,Huber:2020xny,Kalin:2020lmz,Cheung:2020sdj,Cheung:2020gbf,Bern:2020uwk} considered tidal deformations of the astrophysical objects. In the present paper we explore a different direction and focus on effects due to the spin of the compact objects.

When considering intrinsic angular momentum in the problem of a binary of compact astrophysical objects, one assumes that the spin of the objects is subdominant to the angular momentum of the system.  In this way, we organize the effects we consider in a systematic expansion in the spin of the objects. Along these lines, there has been great progress in incorporating spin effects in the PN approximation. Refs.~\cite{Faye:2006gx,Blanchet:2006gy,Damour:2007nc,Steinhoff:2007mb,Steinhoff:2008zr} approached these effects with traditional GR techniques.  Ref.~\cite{Porto:2005ac} extended the worldline EFT methods of Ref. ~\cite{Goldberger:2004jt} in this direction, and since then there have been substantial developments~\cite{Porto:2006bt,Porto:2007qi,Porto:2008jj,Levi:2008nh,Porto:2008tb,Porto:2010tr,Porto:2010zg,Levi:2010zu,Levi:2011eq,Levi:2014sba,Levi:2015uxa,Levi:2015ixa,
Maia:2017gxn,Maia:2017yok,Goldberger:2020fot} (see Refs.~ \cite{Porto:2016pyg,Levi:2018nxp} for reviews).  

The current state-of-the-art results at the 5PN\footnote{The $n$PN order corresponds to $\mathcal{O}(G^a v^{2b}S^c)$ with $a+b+c=n+1$, where $v$ is the relative velocity of the binary system and $S$ corresponds collectively to the spins of the objects. } order include
the linear-in-spin~\cite{Levi:2020kvb} and quadratic-in-spin~\cite{Levi:2020uwu} interactions at next-to-next-to-next-to-leading order
and the cubic-in-spin~\cite{Levi:2019kgk} and quartic-in-spin~\cite{Levi:2020lfn} interactions at next-to-leading order.
The PM literature on the other hand is less developed. Refs.~\cite{Bini:2017xzy,Bini:2018aps} recently obtained results at the 1PM and 2PM orders for effects linear in the spin of the objects via GR considerations. Ref.~\cite{Vines:2017hyw} treated the black hole (BH) case at 1PM order and exactly in the spin by matching an effective action to the linearized Kerr solution. Refs.~\cite{Vines:2018gqi,Siemonsen:2019dsu} obtained the 2PM-order scattering angle in the special kinematic configuration where the spins of the BHs are aligned to the orbital angular momentum of the binary.

Similarly to the non-spinning case, we may use scattering amplitudes to study the gravitational potential between spinning objects. Indeed, Ref. ~\cite{Holstein:2008sx} calculated a one-loop amplitude using Feynman rules, which allowed them to obtain a 2PM-order potential  by means of a  Born iteration.  Following the approach of ~\cite{Neill:2013wsa},  Ref.~\cite{Vaidya:2014kza} reproduced Hamiltonians describing the interactions between spinning BHs by considering spinning particles minimally coupled to gravity. 
Later, Ref.~\cite{Guevara:2017csg} used the generalization of minimal-coupling amplitudes of~\cite{Arkani-Hamed:2017jhn}  and the holomorphic classical limit of~\cite{Cachazo:2017jef} to show that amplitudes encode information about BHs that is exact in spin. 
Refs.~\cite{Guevara:2018wpp, Chung:2018kqs} used the massive spinor-helicity formalism of~\cite{Arkani-Hamed:2017jhn}  to study 2PM-order gravitational scattering from a one-loop amplitude.
Furthermore,  Ref.~\cite{Maybee:2019jus} related classical observables of a scattering process between  spinning particles directly to the scattering amplitude, extending the formalism of~\cite{Kosower:2018adc}. Using this formalism, Refs.~\cite{Chung:2019duq,Chung:2020rrz,Guevara:2019fsj} obtained a 1PM-order Hamiltonian that reproduced the result of~\cite{Vines:2017hyw}. 
Finally,  Ref. ~\cite{Bern:2020buy} obtained the conservative 2PM-order potential that is bilinear in the spin of the objects and valid for arbitrary spin orientations.

Studies of the classical physics of spinning particles have also revealed double copy structures. Refs.~\cite{Arkani-Hamed:2019ymq,Bautista:2019tdr,Chung:2019yfs,Guevara:2020xjx,Emond:2020lwi,Aoude:2020mlg} applied the definition of minimal coupling of ~\cite{Arkani-Hamed:2017jhn} to classical solutions. In this way they made contact with the classical double copy of Ref. ~\cite{Monteiro:2014cda} and with an effective theory of on-shell heavy spinning particles~\cite{Aoude:2020onz}. The latter generalizes the heavy black hole effective theory of Ref.~\cite{Damgaard:2019lfh},  whose amplitudes are known to double copy~\cite{Haddad:2020tvs}.

A suprising structure that emerged from the calculation of Ref.~\cite{Bern:2020buy} is the expression of the observables in a scattering event in terms of the eikonal phase~\cite{Amati:1990xe}. Similar relations already existed in the non-spinning case~\cite{Amati:1990xe,Melville:2013qca,Luna:2016idw,Akhoury:2013yua,KoemansCollado:2019ggb,Cristofoli:2020uzm,DiVecchia:2019myk,DiVecchia:2019kta,Bern:2020gjj,Parra-Martinez:2020dzs}. In the spinning case there was evidence for such a relation in the special kinematic configuration where the spins of the particles are parallel to the angular momentum of the system~\cite{Guevara:2018wpp,Vines:2018gqi,Siemonsen:2019dsu}.  The formula of~\cite{Bern:2020buy} was the first example of such a relation for arbitrary orientations of the spins.  This striking observation potentially implies that all physical observables are obtainable via simple manipulations of the scattering amplitude.

The goal of the present paper is to obtain a 2PM-order Hamiltonian that describes the dynamics between a binary of generic spinning objects  in GR including effects that are up to quadratic in the spin.  We take the masses of the two objects to be $m_1$ and $m_2$ and the rest-frame spin three vectors to be $\bm S_1$ and $\bm S_2$. We denote the relative distance between the objects as $\bm r$ and the momentum three vector in the center-of-mass frame as $\bm p$. The Hamiltonian then reads
\begin{align}
H &=  \null   \sqrt{\bm p ^2 + m_1^2}+\sqrt{\bm p^2 + m_2^2}+V^{(0)}(\bm r^2, \bm p^2)+V^{(1,1)}(\bm r^2, \bm p^2)\frac{ \bm L\cdot \bm S_1}{\bm r^2}
+V^{(1,2)}(\bm r^2, \bm p^2)\frac{ \bm L\cdot \bm S_2}{\bm r^2}
\nn \\
&
+ V^{(2,1)}(\bm r^2, \bm p^2) \frac{(\bm r\cdot \bm S_1 )(\bm r\cdot \bm S_2 ) }{\bm r^4}
+ V^{(2,2)}(\bm r^2, \bm p^2) \frac{\bm S_1 \cdot \bm S_2}{\bm r^2}
+ V^{(2,3)}(\bm r^2, \bm p^2) \frac{( \bm p\cdot \bm S_1 )( \bm p\cdot \bm S_2 )}{\bm r^2}
\nn \\
&
+ V^{(2,4)}(\bm r^2, \bm p^2) \frac{(\bm r\cdot \bm S_1 )^2 }{\bm r^4}
+ V^{(2,5)}(\bm r^2, \bm p^2) \frac{\bm S_1^2}{\bm r^2}
+ V^{(2,6)}(\bm r^2, \bm p^2) \frac{( \bm p\cdot \bm S_1 )^2}{\bm r^2}
%\nn \\
%&
%+ V^{(2,7)}(\bm r^2, \bm p^2) \frac{(\bm r\cdot \bm S_2 )^2 }{\bm r^4}
%+ V^{(2,8)}(\bm r^2, \bm p^2) \frac{\bm S_2^2}{\bm r^2}
%+ V^{(2,9)}(\bm r^2, \bm p^2) \frac{( \bm p\cdot \bm S_2 )^2}{\bm r^2} 
+ \dots \,,
\label{intro_Hamiltonian_general}
\end{align}
where $\bm L = \bm r\times \bm p$ is the orbital angular momentum, and the ellipsis stands for terms of higher order in the spin.  Note that we omit terms quadratic in $\bm S_2$ as they are obtained from the ones quadratic in $\bm S_1$ via appropriate relabeling.
The terms in Eq.~(\ref{intro_Hamiltonian_general}) take the form 
\begin{align}
V^{A}(\bm r^2, \bm p^2)&=\frac{G}{|\bm r|} c_1^{A}(\bm p^2)+ \left(\frac{G}{|\bm r|} \right)^2 c_2^{A}(\bm p^2)+ {\cal O}(G^3) \,, 
\label{intro_potentials}
\end{align}
where the label $A$ takes the values indicated in Eq.~(\ref{intro_Hamiltonian_general}).

Our task is to determine the coefficients $c_i^A$ appearing in Eq. (\ref{intro_potentials}).
For simplicity, and since the bilinear-in-spin interactions were given in Ref.~\cite{Bern:2020buy}, we may consider one of the bodies to be non spinning.  This amounts to formally setting $\bm S_2 = 0$ in Eq.~(\ref{intro_Hamiltonian_general}). We have explicitely verified that the results of this paper do not change if we take into account all the terms in Eq.~(\ref{intro_Hamiltonian_general}). 

Following Refs.~\cite{Cheung:2018wkq,Bern:2020buy}, we obtain the  potential coefficients in question via a matching calculation.
First, we calculate a one-loop scattering amplitude in our so-called full theory. This is a theory that describes particles of arbitrary spin coupled to gravity. Specifically, it captures minimal and non-minimal coupling of the particles to gravity. In terms of our Lagrangian, we include all possible operators that are up to quadratic in the spin of the massive particle and up to linear in the curvature.
Then,  we calculate the corresponding amplitude in an EFT of spinning particles interacting via the Hamiltonian of Eq.~(\ref{intro_Hamiltonian_general}). Our EFT generalizes that of Refs.~\cite{Cheung:2018wkq,Bern:2020buy} to consider effects quadratic in the spin of one of the particles.  

In obtaining these amplitudes we restrict to the piece that captures the classical dynamics. We implement the classical limit by rescaling $q \rightarrow \lambda q, \, S_{1} \rightarrow (1/\lambda)S_{1}$ and expanding in $\lambda$, where $q$ denotes graviton momenta and $S_1$ the covariant spin of the spinning particle.
Finally, we fix the desired coefficients by matching
the two computed amplitudes.

The remainder of this paper is structured as follows:
In Sec. \ref{sec:review}  we review some aspects of the spin formalism introduced in~\cite{Bern:2020buy} that we use throughout the paper.
Namely, we describe our field-theory approach to higher spin and its classical limit. 
We compute the necessary full-theory tree and one-loop amplitudes in Sec.  \ref{sec:full}.
We adopt the method of generalized unitarity~\cite{Bern:1996je,Bern:2011qt,Britto:2004nc} to produce the loop-level amplitude, using tree-level amplitudes as building blocks. 
We then express the amplitudes in the center-of-mass frame,  which facilitates the matching to the EFT.
Sec. \ref{sec:EFT} contains the setup of the EFT, along with the computation of the EFT amplitudes.  
By equating the full-theory and EFT amplitudes, we obtain the desired two-body Hamiltonian. We compare our result against PN~\cite{Levi:2016ofk} and test-body~\cite{Vines:2016unv} Hamiltonians in the literature.
Finally,  in Sec.  \ref{sec:angle} we use the derived Hamiltonian to compute scattering observables.
We then 
establish that the conjecture of Ref.~\cite{Bern:2020buy}, which directly relates these observables to the eikonal phase~\cite{Amati:1990xe}, holds unaltered when we include the quadratic-in-spin effects.
We present our concluding remarks in Sec.~\ref{sec:conclusions}.

\textbf{Note added}: As this paper was in its latest stages we learned about~\cite{PortoPMspin}, which contains overlap with our work. Ref.~\cite{PortoPMspin} extended the worldline PM EFT of~\cite{Kalin:2020mvi,Kalin:2020fhe} to include spin degrees of freedom. We have explicitly verified that, where overlapping,  our results are in agreement with those of~\cite{PortoPMspin}.

\section{Review of Spin Formalism}
\label{sec:review}

%%%%%%%%%%%%%% FIGURE %%%%%%%%%%%%%
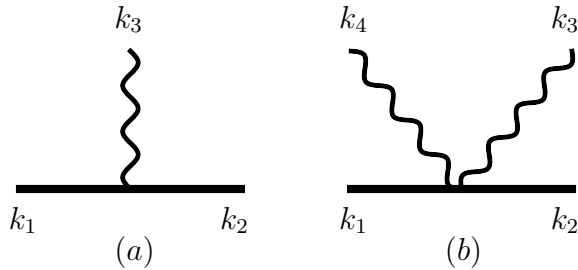
\begin{figure}[tb]
\begin{center}
  \includegraphics[scale=0.65]{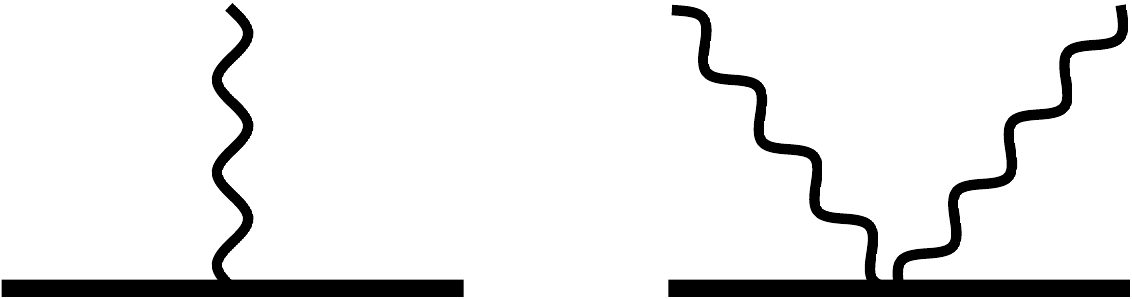}
  \put(-10,-13){$k_2$}\put(-90,-13){$k_1$}
  \put(-135,-13){$k_2$}\put(-215,-13){$k_1$}
  \put(-10,63){$k_3$}\put(-90,63){$k_4$}
  \put(-175,63){$k_3$}
  \put(-175,-25){$(a)$}
  \put(-50,-25){$(b)$}
\end{center}
\vskip -.4 cm
\caption{\small The Feynman vertices used to compute full-theory amplitudes.  The three-particle vertex (a) determines the $\cO(G)$ dynamics.  The Compton amplitude, which requires the contact vertex (b), captures the $\cO(G^2)$ dynamics. The straight lines correspond to the spinning particle, while the wiggly lines correspond to gravitons.}
\label{Vertices}
\end{figure}
%%%%%%%%%%%%%%%%%%%%%%%%%%%%%%%%%%%
In this section, we review the aspects of the higher-spin formalism that we use in the paper. For further details, we refer the reader to Ref.~\cite{Bern:2020buy}.

We identify spinning compact astrophysical objects with higher-spin particles. We describe these massive particles of integer-spin $s$ by real symmetric traceless rank-$s$ tensor fields $\phi_s$. For brevity, we suppress the indices of $\phi_s$, implying matrix multiplication when necessary.

We use a Lagrangian to organize the interactions of higher-spin fields with gravity.  Ref. \cite{Singh:1974qz} obtained such a Lagrangian using auxiliary fields to eliminate all but the spin-$s$ representation of the $SO(3)$ rotation group. Here we relax this requirement, and interpret the theory as a relativistic effective theory that captures all spin-induced multipole moments of spinning objects coupled to gravity.  We write the higher-spin Lagrangian ${\cal L}$ and action $S$ as
\begin{equation}
{\cal L}={\cal L}_{\rm min} + {\cal L}_{\rm nonmin} \ , \qquad S = \int d^4 x \sqrt{-g} {\cal L} \ .
\end{equation}
The minimal Lagrangian contains terms with up to two derivatives,
\begin{align}
{\cal L}_{\rm min} &=-R(e, \omega) + \frac{1}{2} g^{\mu\nu} \nabla(\omega)_\mu\phi_s\nabla(\omega)_\nu\phi_{s} 
- \frac{1}{2}m^2\phi_s\phi_{s}\,.
\label{Ls}
\end{align}
The covariant derivative is
\begin{equation}
 \nabla(\omega)_{\mu} \phi_s \equiv  \partial_{\mu} \phi_s +\frac{i}{2} \omega{}_\mu{}_{ef} M^{ef}\phi_s \, ,
\end{equation}
where $\omega$ is the spin connection, and $M^{ab}$ are the Hermitian Lorentz generators. The gravitational field is described in the vielbein formulation.
The non-minimal Lagrangian containing all the terms linear in the graviton and bilinear in the higher-spin field is
\begin{align}
\label{Lnonmin}
{\cal L}_\text{non-min}&=\sum_{n=1}^{\infty} \frac{\left(-1\right)^n}{\left(2n\right)!}
\frac{C_{ES^{2n}}}{m^{2n}} \nabla(\omega)_{f_{2n}}\cdots \nabla(\omega)_{f_3} R_{f_1 a f_2 b} 
\nabla(\omega)^a{\phi}_s\,  \S^{(f_1} \dots \S^{f_{2n})} \nabla(\omega)^b \phi_s
\\
&-\sum_{n=1}^{\infty} \frac{\left(-1\right)^n}{\left(2n+1\right)!}
\frac{C_{BS^{2n}}}{m^{2n+1}} \nabla(\omega)_{f_{2n+1}}\cdots \nabla(\omega)_{f_3}
\frac{1}{2}\epsilon_{a b (c| f_1} R^{a b}{}_{|d) \, f_2} \nabla(\omega)^c  {\phi}_s  \S^{(f_1} \dots \S^{f_{2n+1})}  \nabla(\omega)^d\phi_s
\,.
\nonumber
\end{align}
where we use an off-shell analog of the Pauli-Lubanski vector 
\begin{equation}
\S^{a}\equiv -\frac{i}{2m}\epsilon^{a b cd}M_{cd}\nabla(\omega)_{b}
\,.
\label{Shat_operator}
\end{equation}
The operators in Eq. (\ref{Lnonmin}) are in direct correspondence to the non-minimal couplings in the worldline spinning-particle action of Ref.  \cite{Levi:2015msa}.  One could, in principle,  include terms with dependence on higher powers of the curvature, but we do not attempt to do so in the present paper. 
Since our objective is to describe the dynamics up to spin squared, we  focus on the first non-minimally coupled term,
\begin{align}
\label{LES2}
{\cal L}_{ES^{2}}=&-\frac{C_{ES^{2}}}{2m^2}
 R_{f_1 a f_2 b} 
\nabla^a{\phi}_s\,  \S^{(f_1} \S^{f_{2})} \nabla^b \phi_s.
\end{align}
Ref.~\cite{Porto:2005ac} first studied the effects captured by this operator at leading order in the PN approximation. The extensions to next-to-leading and next-to-next-to-leading orders were considered in Refs.~\cite{Porto:2008jj} and~\cite{Levi:2016ofk} respectively,  while Ref. ~\cite{Levi:2014gsa} studied its contributions to higher orders in spin.  We instead consider its effects in the PM approximation. 

To extract Feynman rules, we define the graviton as the fluctuation of the metric around Minkowski space.  We determine the spin connection $\omega$ as the solution of the vielbein postulate,  $\nabla_\mu (\omega) e_\nu^{\ a}=0$.  This yields the following expansions for the needed quantities
\begin{align}
& g_{\mu\nu}  = \eta_{\mu\nu}+h_{\mu\nu}\,, \hskip 1. cm 
e{}_\mu{}^a =\delta_\mu^a + \frac{1}{2}h{}_\mu{}^a-\frac{1}{8}h_{\mu\rho}h^{a\rho}+\Ord(h^3) \, ,\nn \\
& \omega(e){}_\mu{}_{cb} = -\partial_{[c}h_{b]\mu}
 - \frac{1}{4} h{}^\rho{}_{[c} \partial_\mu  h{}_{b]\rho}
+ \frac{1}{2} h{}^\rho{}_{[c} \partial_\rho  h{}_{b]\mu}
 - \frac{1}{2} h{}^\rho{}_{[c} \partial_{b]}   h_{\mu\rho} + \Ord(h^3) \, .
\end{align}
After substituting this expansion into the Lagrangian of Eq. (\ref{LES2}),  we follow a straightforward procedure to obtain the Feynman vertices in Fig.  \ref{Vertices}. These are the vertices necessary to determine the dynamics through $\cO(G^2)$. 

We describe the state of the higher-spin particles by their momentum $p$ and polarization tensor $\varepsilon(p)$.  To take the classical limit of expectation values, we choose ``spin coherent states'' \cite{CoherentStates}, whose defining property is that they minimize the standard deviation of observables. Following \cite{Berestetsky:1982aq,Khriplovich:1997ni}, we relate the classical spin tensor and Lorentz generators via
\begin{align}
\pol(\tilde{p})M^{\mu_1\nu_1}\pol(p)
&=S(p)^{\mu_1\nu_1}\pol(\tilde{p})\cdot \pol(p) +\ldots\, , \nonumber \\
\pol(\tilde{p})\{ M^{\mu_1\nu_1},M^{\mu_2\nu_2}\}\pol(p)
&=S(p)^{\mu_1\nu_1} S(p)^{\mu_2\nu_2}\pol(\tilde{p})\cdot \pol(p) +\ldots \, ,
\end{align}
where $\{A,B\} \equiv \frac{1}{2} (AB+BA)$ and $\tilde{p}\equiv -p-q$ (note that we use the all-outgoing convention).  We can also write analogous expressions for products with higher powers of the Lorentz generator.  
Throughout the paper we omit terms that do not contribute to the classical potential  in ellipsis.  These include terms that do not survive in the classical limit and terms that cancel in the matching between full-theory and EFT amplitudes.

Importantly, one can only interpret the symmetric product of Lorentz generators as a product of spin tensors.  However, it is always possible to decompose a product of Lorentz generators into a sum of completely symmetric products by means of the Lorentz algebra,
\begin{equation}
[M^{\mu_1\nu_1},M^{\mu_2\nu_2}]
=i(
\eta^{\mu_3 \mu_1}M^{\mu_4 \mu_2}
+\eta^{\mu_2 \mu_3}M^{\mu_1 \mu_4}
-\eta^{\mu_4 \mu_1}M^{\mu_3 \mu_2}
-\eta^{\mu_2 \mu_4}M^{\mu_1 \mu_3}) \,.
\end{equation}

We take the spin tensor to obey the so-called covariant spin supplementary condition,
\begin{equation}
p_\mu { S}(p)^{\mu \nu} = 0  \, .
\label{CovariantSpinSupplementary}
\end{equation}
We define the Pauli-Lubanski spin vector by
\begin{equation}
S^\alpha(p) = -\frac{1}{2m} \epsilon^{\alpha\beta\gamma\delta}{p}_{\beta}{S}_{\gamma\delta}(p)\,.
\end{equation}
Using the on-shell condition for the spinning particle $p^2 = m^2$ and Eq.~(\ref{CovariantSpinSupplementary}), we find
\begin{equation}
S^{\alpha\beta}(p)= -\frac{1}{m}\epsilon^{\alpha\beta\gamma\delta}{p}_{\gamma}{S}_{\delta}(p)\,.
\end{equation}
For this choice of the spin vector we have
\begin{align}
{ S}(p)^\mu &= \bigg(\, \frac{{\bm p}  \cdot {\bm S}}{m}, {\bm S} + \frac{{\bm p} \cdot {\bm S}}{m(E+m)} {\bm p} \, \bigg)
%\hskip 1.8 cm 
%p_\mu {\rm S}(p, {\bm S})^\mu = 0 
\, ,
\label{covSpinVector}
\end{align}
where ${\bm S}$ is the three-dimensional rest-frame spin of the particle and $p=-(E,{\bm p})$.  I.e. we obtain the covariant spin vector by boosting its rest-frame counterpart.
Finally,  by writing the polarization tensors as boosts of rest-frame coherent
states \cite{CoherentStates}, we have
\begin{align}
\pol(\tilde{p}) \cdot \pol(p) 
 = \exp\left[- \frac{\bm L_q \cdot \bm S}{m(E+m)} \right] +\ldots \, ,
 \label{full_epsdoteps}
\end{align}
where $\bm L_q \equiv i \bm p \times \bm q$, 
and the ellipsis stand for terms that do not contribute to the classical potential.

%%%%%%%%%%%%%% FIGURE %%%%%%%%%%%%%
\begin{figure}[tb]
\begin{center}
  \includegraphics[scale=.65]{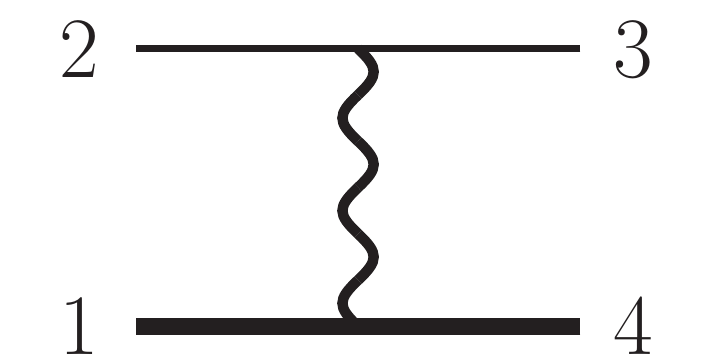}
  %\put(5,-10){$p_4$} \put(-100,-10){$p_1$}
 % \put(5,62){$p_3$} \put(-100,62){$p_2$}
\end{center}
\vskip -.4 cm
\caption{\small The tree-level amplitude that captures the $\Ord(G)$ spin interactions.  The thick (thin) straight line represents the spinning (scalar) particle, while the wiggly line corresponds to the exchanged graviton.}
\label{FourPtTreeFigure}
\end{figure}
%%%%%%%%%%%%%%%%%%%%%%%%%%%%%%%%%%%

\section{Full theory amplitudes}
\label{sec:full}
In this section we calculate the scattering amplitudes needed to construct the desired Hamiltonian. Specifically, we obtain the relevant pieces of the tree-level and one-loop two-to-two scattering amplitude between a scalar and a spinning particle. For the tree-level amplitudes we use the Feynman rules derived in the previous section. We use the generalized unitarity method~\cite{Bern:1996je,Bern:2011qt,Britto:2004nc} for the one-loop amplitude.  Anticipating the comparison to the EFT amplitudes, we specialize our results to the center-of-mass frame.

\subsection{Constructing the full-theory amplitudes}

%%%%%%%%%%%%%% FIGURE %%%%%%%%
\begin{figure}
\begin{center}
\includegraphics[scale=.7]{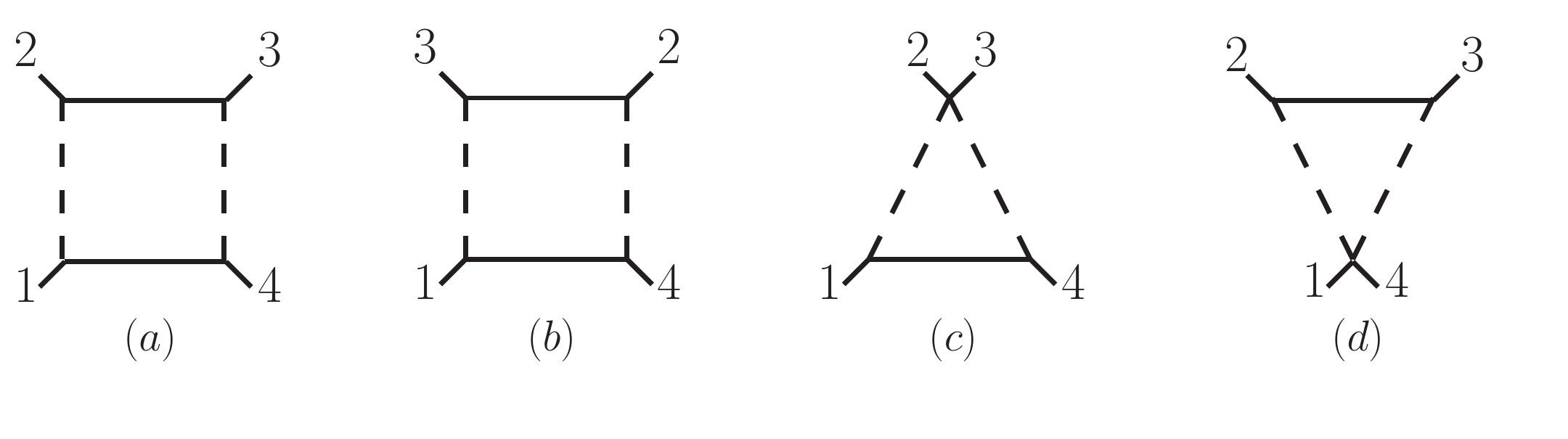}
\end{center}
\vskip -.5 cm
\caption{\small The one-loop scalar box integrals $I_{\Box}$ (a) and $I_{\bowtie}$ (b) and the corresponding triangle integrals $I_{\bigtriangleup}$ (c) and $I_{\bigtriangledown}$ (d). The bottom (top) solid line corresponds to a massive propagator of mass $m_1$ ($m_2$). The dashed lines denote massless propagators.}
\label{OneLoopIntegralsFigure}
\end{figure}
%%%%%%%%%%%%%%%%%%

%\subsection{Tree-level amplitude}

The information to determine the $\mathcal{O}(G)$ Hamiltonian is contained in the tree-level amplitude shown in Fig.~ \Ref{FourPtTreeFigure}.  We take the incoming momentum of the spinning (scalar) particle to be $-p_1$ ($-p_2$) and its outgoing momentum to be $p_4$ ($p_3$). 
Using the Feynman rules obtained above,  we find
\begin{align}
\label{treeS1S2}
\mathcal{M}^{\textrm{tree}} = 
& -\frac{4\pi G}{q^2} \,
\pol_1\cdot \pol_4 \, 
 \left(
\alpha_1^{(0)}
+\alpha_1^{(1,1)} \cE_1
+ 
\alpha_1^{(2,4)}(q\cdot S_1)^2
\right)
+ \ldots\,,
\end{align}
where $\cE_1\equiv i\epsilon^{\mu\nu\rho\sigma}{p_1}_\mu {p_2}_\nu q_\rho S_1{}_\sigma$, and the labeling  scheme for $\alpha_i^A$ follows that for $c_i^A$ in Eq.~(\ref{intro_potentials}). 	In the ellipsis we omit terms that do not contribute to the classical limit, along with pieces proportional to $q^2$, since they cancel the propagator and do not yield long-range contributions. 
The coefficients $\alpha_1^A$ take the explicit form
\begin{align}
\alpha_1^{(0)}=
%%%%% begin : alphaPaper[1,0,1]
4m^4\nu^2(2\sigma^2-1)
%%%%% end : alphaPaper[1,0,1]
\,,\qquad 
\alpha_1^{(1,1)}=
%%%%% begin : alphaPaper[1,1,1]
\frac{8m^2\nu\sigma}{m_{1}}
%%%%% end : alphaPaper[1,1,1]
\,,\qquad 
\alpha_1^{(2,4)}=
%%%%% begin : alphaPaper[1,2,4]
\frac{2 \C2  m^4\nu^2(2\sigma^2-1)}{m_{1}^2}
%%%%% end : alphaPaper[1,2,4]
\,,
\label{eq:treeAlphas}
\end{align}
where we use the variables
\begin{equation}
\sigma= \frac{p_1\cdot p_2}{m_1m_2}\,,\qquad
m=m_1+m_2\,,\qquad \nu=\frac{m_1m_2}{m^2}\,.
\label{eq:varDefs1}
\end{equation}

%\subsection{One-loop amplitude}
In order to construct the $\mathcal{O}(G^2)$ Hamiltonian we further need the corresponding one-loop amplitude. 
We may express any one-loop amplitude as a linear combination of scalar box, triangle,  bubble and tadpole integrals \cite{Passarino:1978jh}. 
Refs.~\cite{Cheung:2018wkq,Bern:2019crd} showed that the bubble and tadpole integrals do not contribute to the classical limit.  Dropping these pieces we may write
\begin{equation}
i\mathcal{M}^\text{1-loop} = d_{\Box}\, I_{\Box} + d_{\bowtie}\, I_{\bowtie}
+ c_{\bigtriangleup} \, I_{\bigtriangleup} + c_{\bigtriangledown}\,  I_{\bigtriangledown} \, ,
\label{Org_Scalar_Integrals}
\end{equation}
where the coefficients $d_{\Box}$, $d_{\bowtie}$,
$c_{\bigtriangleup} $ and $c_{\bigtriangledown} $ are rational
functions of external momenta and polarization tensors. The integrals  $I_{\Box}$, $I_{\bowtie}$,
$I_{\bigtriangleup} $ and $I_{\bigtriangledown} $ are shown in Fig.~\Ref{OneLoopIntegralsFigure}. The triangle integrals take the form~\cite{Cheung:2018wkq}
\begin{eqnarray}
I_{\bigtriangleup,\bigtriangledown}= -\frac{i}{32m_{1,2}} \frac{1}{\sqrt{-q^2}} + \cdots\,.
\label{eq:triangleIntegrals}
\end{eqnarray}
The box contributions do not contain any novel $\mathcal{O}(G^2)$ information. They correspond to infrared-divergent pieces that cancel out when we equate the full-theory and EFT amplitudes~\cite{Cheung:2018wkq,Bern:2019crd}.  In this sense, the explicit values for the box coefficients serve only as a consistency check of our calculation and we do not show them. Instead, we give the result for 
\begin{align}
i \mathcal{M}^{\bigtriangleup+\bigtriangledown}\equiv  c_{\bigtriangleup} \, I_{\bigtriangleup} + c_{\bigtriangledown}\,  I_{\bigtriangledown}\, .
\label{Full_Amp}
\end{align}

\begin{figure}[tb]
\begin{center}
  \includegraphics[scale=.5]{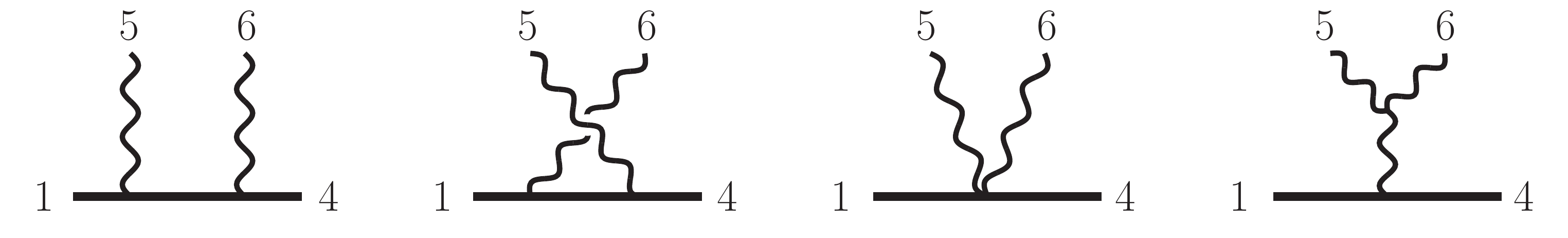}
\end{center}
\vskip -.4 cm
\caption{\small The Compton-amplitude Feynman diagrams. The straight line corresponds to the spinning particle. The wiggly lines correspond to gravitons.  }
\label{fig:ComptonFeynRules}
\end{figure}

\begin{figure}[tb]
\begin{center}
  \includegraphics[scale=.49]{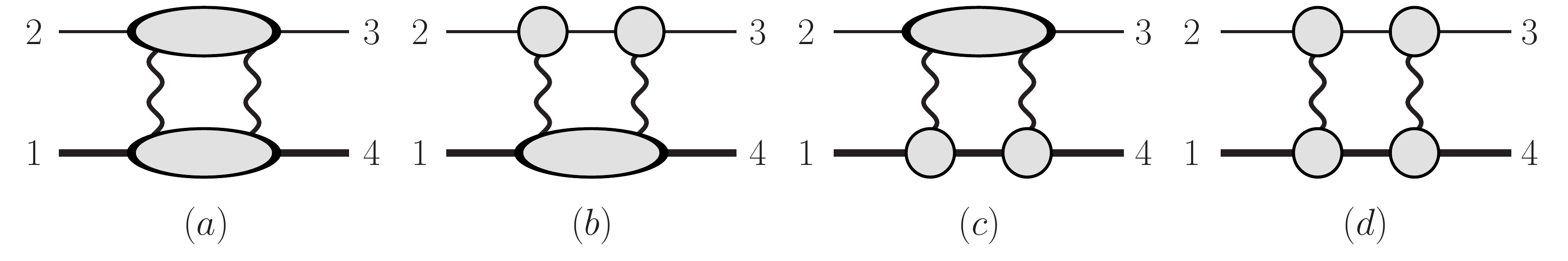}
  \put(-95,50){5} \put(-43,50){6}
  \put(-220,50){5} \put(-168,50){6}
  \put(-345,50){5} \put(-293,50){6}
  \put(-470,50){5} \put(-418,50){6}
\end{center}
\vskip -.4 cm
\caption{\small Appropriate residues of the two-particle cut (a) give the triple cuts (b) and (c), and the quadruple cut (d). The thick straight line corresponds to the spinning particle, the thin straight line to the scalar, and the wiggly lines to the exchanged gravitons. All exposed lines are taken on-shell. }
\label{Cuts}
\end{figure}
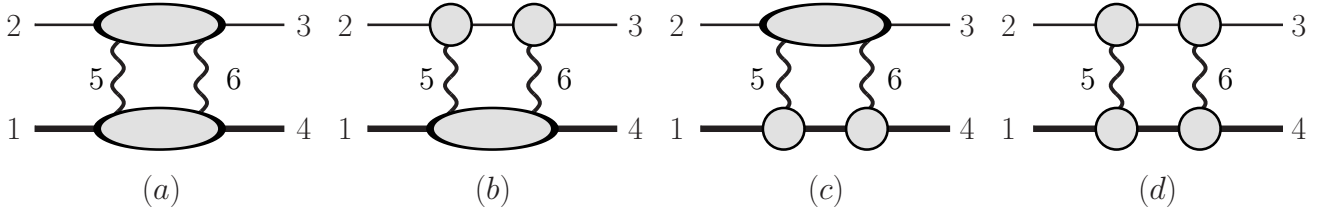

We use the generalized-unitarity method~\cite{Bern:1996je,Bern:2011qt,Britto:2004nc,Ossola:2006us} to obtain the integral coefficients of Eq.~(\Ref{Org_Scalar_Integrals}).  We start by calculating the gravitational Compton amplitude for the spinning particle, using the Feynman rules derived in the previous section.  We depict the relevant Feynman diagrams in Fig.~\Ref{fig:ComptonFeynRules}.  Subsequently, we construct the two-particle cut depicted in Fig.~\Ref{Cuts}(a) by gluing the Compton amplitude for the spinning particle with that for a scalar. The latter is a well-known amplitude.  The residue of the two-particle cut on the scalar-matter pole gives the triple cut in Fig.~\Ref{Cuts}(b), while the one on the spinning-matter pole gives the triple cut in Fig.~\Ref{Cuts}(c). Localizing both matter poles gives the quadruple cut in Fig.~\Ref{Cuts}(d).
Finally, following Refs.~\cite{Forde:2007mi,Kilgore:2007qr,Badger:2008cm},  we obtain the box and triangle coefficients from the quadruple and triple cuts respectively.  Our result reads

%%%%%%%%%%%%%% FIGURE %%%%%%%%%%%%%
%%%%%%%%%%%%%%%%%%%%%%%%%%%%%%%%%%%

%%%%%%%%%%%%%% FIGURE %%%%%%%%%%%%%
%%%%%%%%%%%%%%%%%%%%%%%%%%%%%%%%%%%

\begin{align}
\nonumber
\mathcal{M}^{\bigtriangleup+\bigtriangledown}
 =\frac{2\pi^2G^2 \pol_1\cdot\pol_4}{\sqrt{-q^2}}
\bigg[
\alpha_2^{(0)}+
\alpha_2^{(1,1)}\cE_1+
\alpha_2^{(2,4)}(q\cdot S_1)^2+
\alpha_2^{(2,5)}q^2S_1^2+
\alpha_2^{(2,6)}q^2(p_2\cdot S_1)^2
\bigg]+\ldots\,,
\end{align}
where the coefficients are given by 
\begin{align}
\nonumber
\alpha_2^{(0)}&=
%%%%% begin : alphaPaper[2,0,1]
3m^5\nu^2(5\sigma^2-1)
%%%%% end : alphaPaper[2,0,1]
\,,\qquad 
\alpha_2^{(1,1)}=
%%%%% begin : alphaPaper[2,1,1]
\frac{m^2(4m_1+3m_2)(5\sigma^2-3)\nu\sigma}{m_1(\sigma^2-1)}
%%%%% end : alphaPaper[2,1,1]
\,,
\\
\nonumber 
\alpha_2^{(2,4)}&=
%%%%% begin : alphaPaper[2,2,4]
-\frac{m_2^2}{16(\sigma^2-1)}
\Big[-4m_2\big(-\sigma^2+1+\C2 {} (30\sigma^4-29\sigma^2+3)\big)\\
\nonumber &
\qquad\qquad\qquad\qquad
-m_1\big(35\sigma^4-30\sigma^2-5+\C2 {} (155\sigma^4-174\sigma^2+35)\big)\Big]
%%%%% end : alphaPaper[2,2,4]
\,,\\
\nonumber
\alpha_2^{(2,5)}&=
%%%%% begin : alphaPaper[2,2,5]
-\frac{m_2^2}{16(\sigma^2-1)}
\Big[4m_2\big(15\sigma^4-17\sigma^2+2+\C2 {} (15\sigma^4-13\sigma^2+2)\big)\\
\nonumber &
\qquad\qquad\qquad\qquad
+m_1\big(95\sigma^4-102\sigma^2+7+\C2 {} (95\sigma^4-102\sigma^2+23)\big)\Big]
%%%%% end : alphaPaper[2,2,5]
\,,\\
\nonumber 
\alpha_2^{(2,6)}&=
%%%%% begin : alphaPaper[2,2,6]
-\frac{1}{8(\sigma^2-1)^2}
\Big[2m_2\big(15\sigma^4-14\sigma^2-1+\C2 {} (15\sigma^4-10\sigma^2+3)\big)\\ &
\qquad\qquad\qquad\qquad
+m_1\big(65\sigma^4-66\sigma^2+1+\C2 {} (65\sigma^4-66\sigma^2+17)\big)\Big]
%%%%% end : alphaPaper[2,2,6]
\,.
\label{eq:1loopAlphas}
\end{align}
We note here that the relation $\alpha_2^{(2,4)}=-\alpha_2^{(2,5)}$,  which was expected following a pattern observed in Refs. \cite{Damgaard:2019lfh,Bern:2020buy}, is broken for generic values of $\C2$.  We recover this relation for $\C2=1$, which corresponds to the Kerr black hole \cite{Vines:2017hyw}.  This is in line with a recent observation in Ref.  \cite{Aoude:2020ygw}, that this equality fails to hold in the presence of tidal finite-size effects.

\subsection{The amplitudes in the center-of-mass frame} 

In preparation for the matching procedure in the following section, we specialize our expressions to the center-of-mass frame.  In this frame, the independent four-momenta read
\begin{align}
p_1=-(E_1, \bm p)
\, ,\hskip 1.5 cm 
p_2=-(E_2, -\bm p)
\, , \hskip 1.5 cm 
q=(0, \bm q)
\, , \hskip 1.5 cm 
\bm p\cdot \bm q = \bm q^2/2\,.
\label{COMdef}
\end{align}
Using Eq. (\Ref{covSpinVector}),  we have
\begin{align}
q\cdot S_1 = \bm q\cdot \bm S_1&- \frac{\bm q^2 \bm p\cdot \bm S_1}{2m_1(E_1+m_1)} \,, \qquad
i \epsilon^{\mu\nu\rho\sigma}
{p_1}_\mu {p_2}_\nu q_\rho {S_1}_\sigma = E\,  \bm L_q \cdot \bm S_1 \, , 
\qquad
p_2\cdot S_1 =-\frac{E}{m_1}\vp \cdot \vS1 \,.
\label{VdotS}
\end{align}
Furthermore, Eq.~(\Ref{full_epsdoteps}) becomes
\begin{align}
\pol_1 \cdot \pol_4
 = 1-   \frac{\bm L_q \cdot \bm S_1}{m_1(E_1+m_1)} + \frac{( \bm L_q \cdot \bm S_1 )^2}{2m_1^2(E_1+m_1)^2}+\ldots \, .
 \label{epseps14}
\end{align}
Using the above expressions, our amplitudes take the form
\begin{align}
\label{fullTheoryM}
\frac{\mathcal{M}^\text{tree}}{4E_1E_2}&=
\frac{4 \pi  G }{\bm q^2}
\bigg[
a^{(0)}_{1} 
+a^{(1,1)}_{1} \bm L_q \cdot \bm S_1 
+ a^{(2,4)}_{1} (\bm q\cdot \bm S_{1})^2
\bigg]
\, ,\\
\nonumber 
\frac{\mathcal{M}^{\bigtriangleup+\bigtriangledown}}{4E_1E_2} &=
\frac{2\pi^2  G^2 }{|\bm q |}
\bigg[
a^{(0)}_{2} 
+ a^{(1,1)}_{2} \bm L_q \cdot \bm S_1
+ a^{(2,4)}_{2}  (\bm q\cdot \bm S_{1})^2
+ a^{(2,5)}_{2}  \bm q^2 \, \bm S_1^2\
+ a^{(2,6)}_{2} \bm q^2 \, (\bm p\cdot \bm S_1)^2
\bigg] \, .
\end{align}
The coefficients $a_i^A$ are given in terms of the $\alpha_i^{A}$ of Eqs.~(\Ref{eq:treeAlphas}) and (\Ref{eq:1loopAlphas}) by\footnote{Note that unlike Ref. \cite{Bern:2020buy} we do not introduce the coefficients $a_{cov}$. This means that factors of the spin in Eq.~(\Ref{fullTheoryM}) appear both because we specialize in the center-of-mass frame and due to Eq.~(\Ref{epseps14}). }
\begin{align}
a^{(0)}_i & =  
%%%%% begin : aPaper[i,0,1]
\frac{\alpha_i^{(0)}}{4m^2\gamma^2\xi}
%%%%% end : aPaper[i,0,1]
\,, \qquad\ 
a^{(1,1)}_i = 
%%%%% begin : aPaper[i,1,1]
\frac{\alpha_i^{(1,1)}}{4m\gamma\xi}-\frac{1}{m_1^2(\gamma_1+1)} \frac{\alpha_i^{(0)}}{4m^2\gamma^2\xi}
%%%%% end : aPaper[i,1,1]
\,,  \nn \\
a^{(2,j)}_i & = 
%%%%% begin : aPaper[i,2,j]
\frac{\alpha_i^{(2,j)}\tilde{\zeta}^{(j)}}{4m^2\gamma^2\xi}
-\frac{\zeta^{(j)}}{m_1^2(\gamma_1+1)}\frac{\alpha_i^{(1,1)}}{4m\gamma\xi}
+\frac{\zeta^{(j)}}{m_1^4 (\gamma_1+1)^2} \frac{\alpha_i^{(0)}}{8m^2\gamma^2\xi}
%%%%% end : aPaper[i,2,j]
\, ,
\label{eq:a_covRelations}
\end{align}
where $i=1,2$ and the structure-dependent coefficients are given by
\begin{align}
\zeta^{(4)}=-\zeta^{(5)}=\vp^2, \qquad 
\zeta^{(6)}=1, \qquad
\tilde{\zeta}^{(4)}=\tilde{\zeta}^{(5)}=1, \qquad 
\tilde{\zeta}^{(6)}=-\frac{E^2}{m_1^2}.
\end{align}
In addition to the definitions in Eq.~(\Ref{eq:varDefs1}) we use
\begin{equation}
\gamma = \frac{E}{m} \,,\qquad
\gamma_1 = \frac{E_1}{m_1} \,, \qquad
E = E_1 + E_2 \,, \qquad
\xi = \frac{E_1 E_2}{E^2} \,.
\label{eq:varDefs2}
\end{equation}

\section{Hamiltonian from effective field theory }
\label{sec:EFT}

We now turn our attention to the task of translating the scattering amplitudes of higher-spin fields to a two-body conservative Hamiltonian. We do this by matching the scattering amplitude computed in the last section to the two-to-two amplitude of an EFT of the positive-energy modes of higher-spin fields.  Ref.~\cite{Cheung:2018wkq} developed this matching procedure for higher orders in $G$ and all orders in velocity, while Ref.~\cite{Bern:2020buy} extended the formalism to include spin degrees of freedom. We conclude this section by comparing our answer with previous results in the literature.

\subsection{EFT scattering amplitudes}
The action of the effective field theory for the higher-spin fields $\xi_{1}$ and $\xi_2$ is given by
\begin{align}
	S =& \int_{\bm k}   \, 
\sum_{a=1,2} \xi_a^\dagger(-\bm k) \left(i\partial_t - \sqrt{\bm k^2 + m_a^2}\right) \xi_a(\bm k)
-\int_{\bm k, \bm k'} \, 
 \xi_1^\dagger(\bm k') \xi_2^\dagger(-\bm k') \,
  \Vmom(\bm k' ,\bm k, \pS_1) \,\xi_1(\bm k)\xi_2(-\bm k) \, ,
\label{eq:eftL}
\end{align}
where $\int_{\bm k} = \int \frac {d^{D-1} \bm k}{(2\pi)^{D-1}}$,  and the interaction potential $\Vmom(\bm k' ,\bm k, \pS_1)$ is a function 
of the incoming and outgoing momenta $\bm k$ and $\bm k'$, and the spin operator
$\pS_1$.  
We consider kinematics in the center-of-mass frame. As in the full theory side,  we choose the field $\xi_2$ to be a scalar,  while the asymptotic states of $\xi_1$ are taken to be spin coherent states. We obtain the classical rest-frame spin vector as the expectation value of the spin operator with respect to these on-shell states.  

We build the most general potential containing only long-range classical contributions, up to quadratic order in spin.  In momentum space, a minimal basis of interactions in the on-shell scheme is given by the operators
%%%
\begin{equation}
\label{OperatorList}
\hat{\mathbb{O}}^{(0)}=\mathbb{I} \,,  
\hspace{.5cm} 
\hat{\mathbb{O}}^{(1,1)}= \FTLQ\cdot \pS_1 \,, 
\hspace{.5cm} 
\hat{\mathbb{O}}^{(2,4)} = \big( \hat{\bm{q}} \cdot \bm{\hat{S}}_1 \big)^2 \,, 
\hspace{.5cm} 
\hat{\mathbb{O}}^{(2,5)} = \hat{\bm{q}}^2\ \bm{\hat{S}}_1^2  \,, 
\hspace{.5cm} 
\hat{\mathbb{O}}^{(2,6)} = \hat{\bm{q}}^2 \big( \bm{k} \cdot \bm{\hat{S}}_1 \big)^2 \,,
\end{equation}
where\footnote{The three-vectors $\hat{\bm q}$ and $\hat{\bm p}$ are not to be confused with unit-norm vectors.} $\hat{\bm{q}} \equiv \bm k - \bm k^\prime$ and $\FTLQ \equiv i \bm k \times \bm \hat{\bm{q}} $.
Their expectation values with respect to spin coherent states are in one-to-one correspondence with the monomials in the full theory amplitude, Eq. (\Ref{fullTheoryM}). The labeling scheme for the operators follows the conventions of Eq.~(\ref{intro_Hamiltonian_general}).  We use the following ansatz for the potential operator
\begin{align}
\Vmom(\bm k', \bm k, \pS_1) =& 
\sum_A \Vmom^{A}(\bm k', \bm k)\, \hat \Operator^A \,,
\label{eq:V_mom}
\end{align}
where $A$ runs over the superscripts of the operators in \eqn{OperatorList}.
 $\Vmom^{A}(\bm k',\bm k)$ are free coefficients 
with the same structure as the spin-independent potential of Refs. \cite{Bern:2019crd,Bern:2019nnu},
\begin{align}
\Vmom^{A}(\bm k',\bm k) =& \frac{4\pi G}{\hat{ \bm q}^2} d^A_1\left(\hat{\bm p}^2 \right)
+ \frac{2\pi^2 G^2}{|\hat{\bm q}|}  d^A_2\left(\hat{\bm p}^2 \right)+  \Ord(G^3) \,,
\label{eq:V_p}
\end{align}
where $\hat{\bm p}^2 \equiv ( \bm k^2 + \bm k^{\prime 2} )/2$.
At the $\mathcal{O}(G)$ level, the operators containing a factor of $\hat{ \bm q}^2$ can be ignored, as they lead to contact terms. Therefore we choose
\begin{equation}
d^{(2,5)}_1 = d^{(2,6)}_1 =0 \,.
\end{equation}
However, the  factor of $\hat{ \bm q}^2$ does not cancel out with the $\mathcal{O}(G^2)$ denominator, so we need to keep $d^{(2,5)}_2$ and $d^{(2,6)}_2$. 

We now evaluate the EFT two-to-two scattering amplitude.  To this end we use the Feynman rules derived from the EFT action (Eq. (\Ref{eq:eftL})),
\begin{equation}
	\includegraphics[scale=.55,trim={0 0.2cm 0 0}, clip]{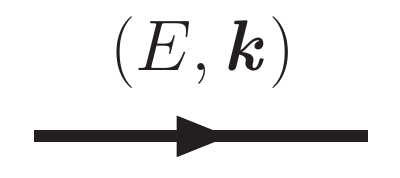} = \frac{i\, \mathbb{I}}{E-\sqrt{\bm k^2+m^2}+i\epsilon}\,, \qquad
	\vcenter{\hbox{\includegraphics[scale=.55,trim={0 0.5cm 1cm 0}, clip]{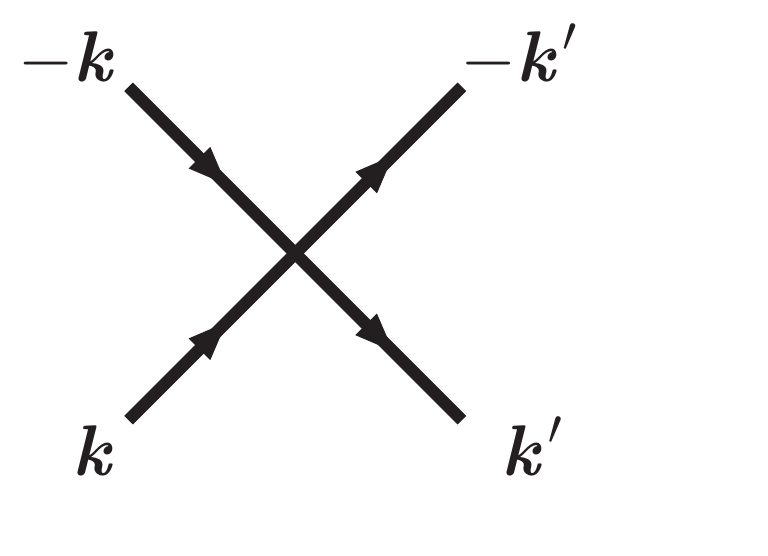}}} =
	-i V(\bm k', \bm k, \pS_{1})\,.
\label{eq:EFT_FeynmanRules}
\end{equation}
Using these rules we compute the amplitude up to $\cO(G^2)$ directly evaluating the relevant Feynman diagrams, omitting terms that do not contribute to long range interactions. The spin-dependent vertices must be treated as operators,  and thus their ordering is important.  After carrying out the energy integration,  we obtain an expression  for the amplitude
\begin{align}
\mathcal{M}^{\textrm{EFT}}
=& -\Vmom(\bm p',\bm p, \bm S_1) - \int_{\bm k} \frac{\Vmom(\bm p',\bm k, \bm S_1)\, \Vmom(\bm k,\bm p, \bm S_1)}{E_1+E_2- \sqrt{\bm k^2+m_1^2}- \sqrt{\bm k^2+m_2^2}} \ .
\label{eq:eft_FeynmanDiagram}
\end{align}
Similarly to the full theory, in order to extract the classical limit, one needs to first decompose products of the spin vector into irreducible representations of the rotation group, by repeated use of the $SO(3)$ algebra.

At $\cO(G)$ the EFT amplitude receives a contribution only from the first term of Eq.~(\Ref{eq:eft_FeynmanDiagram}),
\begin{align}
\mathcal{M}^{\rm EFT}_{\rm 1PM} =
\frac{4\pi G }{\bm q^2}
\left[
a^{(0)}_1 + a^{(1,1)}_1 \FTLq\cdot \bm S_1 
+ a^{(2,4)}_1 \, (\bm q\cdot \bm S_1)^2 \right]\,.
\label{eq:M_1PM}
\end{align}
The $a_1^A$ are given directly in terms of the momentum-space potential coefficients,
\begin{equation}
a^A_1 = -d^A_1 \,.
\label{eq:a1}
\end{equation}

The EFT amplitude at $\mathcal{O}(G^2)$ receives contributions
from both terms in Eq.  (\Ref{eq:eft_FeynmanDiagram}) and can be
written as
\begin{align}
\mathcal{M}^{\rm EFT}_{\rm 2PM} & =
\frac{2\pi^2 G^2}{|\bm q|}
\bigg[
a^{(0)}_2 + a^{(1,1)}_2 \FTLq\cdot \bm S_1
+ a^{(2,4)}_2 (\bm q\cdot \bm S_{1})^2
+ a^{(2,5)}_2 \bm q^2 \, \bm S_1^2
+ a^{(2,6)}_2 \bm q^2 \, (\bm p\cdot \bm S_1)^2
\bigg] \nn \\
& \hskip 6 cm \null 
+(4\pi G)^2\, a_{\rm iter}\int \frac{d^{D-1}\bm \ell}{(2\pi)^{D-1}}
\frac{2\xi E}{\bm \ell^2 (\bm \ell+\bm q)^2 (\bm \ell^2+2\bm p\cdot \bm \ell)} \,,
\label{eq:M_2PM}
\end{align}
where $\bm \ell = \bm k - \bm p$ and we only keep terms that are relevant in the 
classical limit.
The above coefficients are given by
\begin{align}
\label{eq:a2}
a^{(0)}_{2} &=
%%%%% begin : aEftPaper[2,0,1]
-d^{(0)}_2+\frac{1}{2\xi E}\, \tilde{A}_0\left[\left(d^{(0)}_1 \right)^2\right] 
%%%%% end : aEftPaper[2,0,1]
\,, \\ \nn
a^{(1,1)}_{2} &=
%%%%% begin : aEftPaper[2,1,1]
-d^{(1,1)}_2+\frac{1}{2\xi E}\, \tilde{A}_2\left[d^{(0)}_1 d^{(1,1)}_1\right]
%%%%% end : aEftPaper[2,1,1]
\,,\\ \nn
a_2^{(2,4)} &= 
%%%%% begin : aEftPaper[2,2,4]
-d_2^{(2,4)}
+ \frac{3}{8 \xi E} \tilde{A}_{4/3} \bigg[ d_1^{(0)} d_1^{(2,4)} \bigg]
+ \frac{\bm{p}^2}{16 \xi E} \tilde{A}_4 \left[ \left(d_1^{(1,1)}\right)^2 \right]
+ \frac{\xi E}{4} d_1^{(1,1)} d_1^{(2,4)}
%%%%% end : aEftPaper[2,2,4]
 \,, \\ \nn
a_2^{(2,5)} &= 
%%%%% begin : aEftPaper[2,2,5]
-d_2^{(2,5)}
- \frac{1}{8 \xi E} \tilde{A}_4 \bigg[ d_1^{(0)} d_1^{(2,4)} \bigg]
- \frac{\bm{p}^2}{8 \xi E} \tilde{A}_4 \bigg[ \left(d_1^{(1,1)}\right)^2 \bigg]
+ \frac{\xi E}{4} d_1^{(1,1)} d_1^{(2,4)} 
%%%%% end : aEftPaper[2,2,5]
\,, \\ \nn
a_2^{(2,6)} &= 
%%%%% begin : aEftPaper[2,2,6]
-d_2^{(2,6)}
+ \frac{\xi E}{\bm{p}^4} d_1^{(0)} d_1^{(2,4)}
+ \frac{1}{8 \xi E} \tilde{A}_4 \bigg[ \left(d_1^{(1,1)}\right)^2 \bigg]
- \frac{\xi E}{\bm{p}^2} d_1^{(1,1)} d_1^{(2,4)}
%%%%% end : aEftPaper[2,2,6]
 \,,
\end{align}
where we define the function
\begin{align}
\tilde{A}_j[X] &= 
\left[(1-3\xi) + \frac{j \xi^2 E^2}{\bm p^2} + 2 \xi^2 E^2\partial \right] X,
%%%%% end : A1[X]
\end{align}
and the derivative is taken with respect to the square of the center-of-mass momentum $\partial = \partial/\partial{\bm p^2}$. 
The second term in Eq.~(\Ref{eq:M_2PM}) is infrared divergent and should cancel out when we equate the full-theory and EFT amplitudes. We have explicitly verified this cancellation at leading order in the classical expansion.  

\subsection{Conservative spin Hamiltonian}

As mentioned in Sec.  \Ref{sec:intro}, our final result is the position-space Hamiltonian,
\begin{align}
H = \null &  \sqrt{\bm p ^2 + m_1^2}+\sqrt{\bm p^2 + m_2^2}+V^{(0)}(\bm r^2, \bm p^2)+V^{(1,1)}(\bm r^2, \bm p^2)\frac{ \bm L\cdot \bm S_1}{\bm r^2}
\nn \\
&
+ V^{(2,4)}(\bm r^2, \bm p^2) \frac{(\bm r\cdot \bm S_1 )^2 }{\bm r^4}
+ V^{(2,5)}(\bm r^2, \bm p^2) \frac{\bm S_1^2}{\bm r^2}
+ V^{(2,6)}(\bm r^2, \bm p^2) \frac{( \bm p\cdot \bm S_1 )^2}{\bm r^2} + \dots \,.
\label{Hamiltonian_general}
\end{align}
The potentials take the form 
\begin{align}
V^{A}(\bm r^2, \bm p^2)&=\frac{G}{|\bm r|} c_1^{A}(\bm p^2)+ \left(\frac{G}{|\bm r|} \right)^2 c_2^{A}(\bm p^2)+ {\cal O}(G^3) \,.
\end{align}
We obtain the position-space Hamiltonian by taking the Fourier transform of the momentum-space Hamiltonian with respect to the momentum transfer $\vq$, which is the conjugate of the separation between the particles $\bm r$. 
In this way, we express the position-space coefficients $c_i^A$ in terms of the momentum-space coefficients $d_i^A$ via linear relations dictated by the $\bm q$-dependence of the spin operators,
\begin{alignat}{9}
c_1^{(0)} &= d_1^{(0)}\,, \quad
&&c_1^{(1,1)} &&= -d_1^{(1,1)}\,,\quad
&&c_1^{(2,4)} &&= -3d_1^{(2,4)}\,,\quad
&&c_1^{(2,5)} &&= d_1^{(2,4)}\,,\quad
&&c_1^{(2,6)} &&= 0\,,
\label{eq:2PM_potential_real}
\\
\nonumber 
c_2^{(0)} &= d_2^{(0)}\,, \quad
&&c_2^{(1,1)} &&= -2d_2^{(1,1)}\,,\quad
&&c_2^{(2,4)} &&= -8d_2^{(2,4)}\,,\quad
&&c_2^{(2,5)} &&= 2d_2^{(2,4)}-2d_2^{(2,5)}\,,\quad
&&c_2^{(2,6)} &&= -2d_2^{(2,6)} \,.
\end{alignat} 
We determine the momentum-space coefficients $d_i^A$ in terms of the amplitudes coefficients $a_i^A$ by the relations in Eqs. (\Ref{eq:a1}) and (\Ref{eq:a2}). 
We may now obtain $a_i^A$ by demanding that the EFT amplitude matches the full-theory one,
\begin{align}
{\cal M}^\text{EFT}_\text{1PM} = \frac{{\cal M}^\text{tree}}{4E_1 E_2}
\ , \qquad
{\cal M}^\text{EFT}_\text{2PM} = \frac{{\cal M}^\text{1-loop}}{4E_1 E_2}
\ ,
\label{IDamplitudes}
\end{align}
where the factors of the energy account for the non-relativistic normalization of the EFT amplitude. 
Using Eq.  (\Ref{eq:a_covRelations}) we relate $a_i^A$ to $\alpha_i^A$, which are explicitly shown in Eqs.  (\Ref{eq:treeAlphas}) and (\Ref{eq:1loopAlphas}).
Putting everything together, we obtain novel expressions for the position-space coefficients $c_i^A$ which are lengthy, and so we only provide them in the ancillary file coefficients.m. 

%%%%

\subsection{Comparison to the literature}
In order to ensure the validity of our result, we compare it with existing Hamiltonians in the General Relativity literature. Specifically, we compare with overlapping results in Ref.~\cite{Levi:2016ofk}, which obtained the next-to-next-to-leading order post-Newtonian Hamiltonian, and in Ref.~\cite{Vines:2016unv}, which calculated the test-body Hamiltonian. Both references included interactions of up to quadratic order in the spins.

One way to establish the equivalence of two Hamiltonians is to construct a canonical transformation that extrapolates between them. Alternatively, we may compare the gauge invariant scattering amplitudes calculated from the two Hamiltonians by means of the EFT.  We take the latter approach here. To do so, we promote the spin vector  in the classical Hamiltonians to the spin operator, and we account for the non-isotropic terms according to the conventions of \cite{Cheung:2018wkq,Bern:2019nnu,Bern:2019crd}.

In this way we obtain EFT amplitudes in the form of Eqs.~(\Ref{eq:M_1PM}) and (\Ref{eq:M_2PM}). The relevant coefficients for our purposes obtained using the Hamiltonian of Ref. \cite{Levi:2016ofk} read
\begin{align}
%a_1^{(0)}&={m_1} {m_2}+
%\frac{3m_1^2+8m_1m_2+3m_2^2}{2m_1m_2}
%{\vpp} +\frac{18 {m_1}^2 {m_2}^2-5 {m_1}^4-5 {m_2}^4}{8 {m_1}^3 {m_2}^3}{\vp^4}
%\\
%a_1^{(1,1)}&=\frac{4m_1+3 {m_2}}{2 {m_1}}+\frac{18 {m_1}^2+8 {m_1}m_2-5 {m_2^2}}{8 {m_1}^3m_2}{\vpp}
%\\ \nonumber &\qquad\qquad\qquad
%-\frac{15 {m_1}^2 {m_2}^2+15 {m_1}^4+12 {m_1} {m_2}^3-7 {m_2}^4}{16 {m_1}^5 {m_2}^3}{\vp^4} 
%\\
a_1^{(2,4)}&=
%%%%% begin : aLeviPaper[1,2,4]
\frac{m_ 2\C2}{2 m_ 1} - \frac{8 m_ 1 m_ 2 + 7 m_ 2^2-\C2\left(6 m_ 1^2 + 16 m_ 1 m_ 2  + 6 m_ 2^2 \right)}{8 m_ 1^3 m_ 2} {\vp}^2
\\ \nonumber
&\qquad\qquad- \frac{3 m_ 2^2 (7 m_ 1^2 + 4 m_ 1 m_ 2 - 2 m_ 2^2) + \C2(5 m_ 1^4 - 18 m_ 1^2 m_ 2^2 + 5 m_ 2^4)}{16 m_ 1^5 m_ 2^3} {\vp}^4
%%%%% end : aLeviPaper[1,2,4]
+\ldots\,,
\end{align}
and
\begin{align}
%a_2^{(0)}&=3 {m_1} {m_2} ({m_1}+{m_2})+\frac{3({m_1}+{m_2})\left(3 {m_1}^2+10 {m_1} {m_2}+3 {m_2}^2\right) }{4 {m_1} {m_2}} {\vpp} 
%\\
%a_2^{(1,1)}&=\frac{{m_1} {m_2}^2 (4 {m_1}+3 {m_2})}{2 {\vpp} ({m_1}+{m_2})}+\frac{53 {m_1}^2 {m_2}+20 {m_1}^3+41 {m_1} {m_2}^2+9 {m_2}^3}{4 {m_1}^2+4 {m_1} {m_2}}
%\\ \nonumber
%&\qquad\qquad\qquad 
%+\frac{3 \left(43 {m_1}^2 {m_2}^2+71 {m_1}^3 {m_2}+30 {m_1}^4-{m_1} {m_2}^3-4 {m_2}^4\right)}{16 {m_1}^3 {m_2} ({m_1}+{m_2})}{\vpp}\\
\nn a_2^{(2,4)}&=
%%%%% begin : aLeviPaper[2,2,4]
\frac{m_1 m_2^3\C2}{4(m_1+m_2)\vpp}
+\frac{m_2\big(
10m_1^2-7m_1 m_2-13m_2^2
+\C2(32m_1^2+61m_1 m_2+29m_2^2)\big)
}{16m_1(m_1+m_2)}\\& \qquad\qquad\qquad 
\nonumber
+\frac{
15m_1^4-73m_1^3m_2-361m_1^2m_2^2
-343m_1 m_2^3-82m_2^4}
{64m_1^3m_2(m_1+m_2)}\vpp\\& \qquad\qquad\qquad 
 \nonumber
+\frac{
\C2(93m_1^4+467m_1^3m_2+707m_1^2m_2^2
+397m_1 m_2^3+64m_2^4)}
{64m_1^3m_2(m_1+m_2)}\vpp
%%%%% end : aLeviPaper[2,2,4]
+\ldots\,,
\\
a_2^{(2,5)}&=
%%%%% begin : aLeviPaper[2,2,5]
-\frac{m_1 m_2^3\C2}{4(m_1+m_2)\vpp}
-\frac{m_2\big(
22m_1^2+19m_1 m_2+m_2^2
+\C2(20m_1^2+35m_1 m_2+15m_2^2)\big)
}{16m_1(m_1+m_2)}\\& \qquad\qquad\qquad 
\nonumber
-\frac{
51m_1^4+115m_1^3m_2-53m_1^2m_2^2
-155m_1 m_2^3-50m_2^4}
{64m_1^3m_2(m_1+m_2)}\vpp\\& \qquad\qquad\qquad 
\nonumber
-\frac{
\C2(57m_1^4+279m_1^3m_2+399m_1^2m_2^2
+209m_1 m_2^3+32m_2^4)}
{64m_1^3m_2(m_1+m_2)}\vpp
%%%%% end : aLeviPaper[2,2,5]
+\ldots\,,
\\
\nn a_2^{(2,6)}&=
%%%%% begin : aLeviPaper[2,2,6]
\frac{m_1 m_2^3\C2}{2(m_1+m_2)\vp^4}
+\frac{m_2\big(
8m_1^2+8m_1 m_2+m_2^2
+\C2(7m_1^2+11m_1 m_2+4m_2^2)\big)
}{4m_1(m_1+m_2)\vpp}\\& \qquad\qquad\qquad 
\nonumber
+\frac{
33m_1^4+97m_1^3m_2+13m_1^2m_2^2
-73m_1 m_2^3-28m_2^4}
{32m_1^3m_2(m_1+m_2)}\\& \qquad\qquad\qquad 
\nonumber
+\frac{
\C2(39m_1^4+185m_1^3m_2+245m_1^2m_2^2
+115m_1 m_2^3+16m_2^4)}
{32m_1^3m_2(m_1+m_2)}
%%%%% end : aLeviPaper[2,2,6]
+\ldots\,,
\end{align}
where the ellipsis stands for higher orders in $\vp$.  These coefficients are in complete agreement with the velocity expansion of our amplitudes. The Hamiltonian of Ref.~\cite{Vines:2016unv} produces the coefficients
\begin{align}
%a_1^{(0)}&=\frac{\left(2 {\gamma_1}^2-1\right) {m_1} {m_2}}{{\gamma_1}},
%\qquad\qquad\qquad
%a_1^{(1,1)}=\frac{2 {\gamma_1} {m_2}+{m_2}}{{\gamma_1}^2 {m_1}+{\gamma_1} {m_1}}\\
a_1^{(2,4)}&=
%%%%% begin : aVinesPaper[1,2,4]
\frac{{m_2} \left({\C2} \left(2 {\gamma_1}^3+2 {\gamma_1}^2-{\gamma_1}-1\right)-2 {\gamma_1}^3-2 {\gamma_1}^2+3 {\gamma_1}+1\right)}{2 {\gamma_1} ({\gamma_1}+1) {m_1}}
%%%%% end : aVinesPaper[1,2,4]
 \,,
\end{align}
and 
\begin{align}
%a_2^{(0)}&=\frac{3 \left(5 {\gamma_1}^2-1\right) {m_1} {m_2}^2}{4 {\gamma_1}},
%\qquad\qquad\quad
%a_2^{(1,1)}=\frac{3 \left(5 {\gamma_1}^2-2 {\gamma_1}-1\right) {m_2}^2}{4 {\gamma_1} \left({\gamma_1}^2-1\right) {m_1}},
%\\
\nn a_2^{(2,4)}&=
%%%%% begin : aVinesPaper[2,2,4]
\frac{{m_2}^2 \left({\C2} \left(30 {\gamma_1}^4-29 {\gamma_1}^2+3\right)-30 {\gamma_1}^4+59 {\gamma_1}^2-24 {\gamma_1}-5\right)}{16 {\gamma_1} \left({\gamma_1}^2-1\right) {m_1}}
%%%%% end : aVinesPaper[2,2,4]
\,,
\\
\nn a_2^{(2,5)}&=
%%%%% begin : aVinesPaper[2,2,5]
-\frac{{m_2}^2 \left({\C2} \left(15 {\gamma_1}^4-13 {\gamma_1}^2+2\right)-15 {\gamma_1}^4+43 {\gamma_1}^2-24 {\gamma_1}-4\right)}{16 {\gamma_1} \left({\gamma_1}^2-1\right) {m_1}}
%%%%% end : aVinesPaper[2,2,5]
\,,
\\
a_2^{(2,6)}&=
%%%%% begin : aVinesPaper[2,2,6]
\frac{{m_2}^2 \left({\C2} \left(15 {\gamma_1}^4-10 {\gamma_1}^2+3\right)-15 {\gamma_1}^4+46 {\gamma_1}^2-24 {\gamma_1}-7\right)}{16 {\gamma_1} \left({\gamma_1}^2-1\right)^2 {m_1}^3}
%%%%% end : aVinesPaper[2,2,6]
\,,
\\
\nn a_2^{(2,\tilde{4})}&=
%%%%% begin : aVinesPaper[2,2,4,t]
\frac{\left(95 {\gamma_1}^4-102 {\gamma_1}^2+15\right) {m_1}}{32 {\gamma_1} \left({\gamma_1}^2-1\right)}
%%%%% end : aVinesPaper[2,2,4,t]
\,,
\qquad
a_2^{(2,\tilde{5})}=
%%%%% begin : aVinesPaper[2,2,5,t]
\frac{95 {\gamma_1}^4 {m_1}-102 {\gamma_1}^2 {m_1}+15 {m_1}}{32 {\gamma_1}-32 {\gamma_1}^3}
%%%%% end : aVinesPaper[2,2,5,t]
\,,
\\
\nn a_2^{(2,\tilde{6})}&=
%%%%% begin : aVinesPaper[2,2,6,t]
\frac{65 {\gamma_1}^4-66 {\gamma_1}^2+9}{16 {\gamma_1} \left({\gamma_1}^2-1\right)^2 {m_1}}
%%%%% end : aVinesPaper[2,2,6,t]
\,,
\end{align}
where the coefficients $a_2^{(2,\jmath)}$ correspond to the spinning particle as the test body, while $a_2^{(2,\tilde{\jmath})}$ correspond to the scalar particle as the test body.  These coefficients exactly reproduce the test body expansion of our amplitudes. 

\section{Observables from the eikonal phase}
\label{sec:angle}

The conservative Hamiltonian we obtained in the previous section enables the calculation of physical observables for a binary of compact objects interacting through gravity.  
On the one hand, one may calculate quantities that describe bound trajectories of the binary,  as the bound-state energy. 
On the other hand, observables pertaining to unbound orbits have received a surge of attention. 
The main reason for this is that these observables serve as input to important phenomenological models, as the effective one-body Hamiltonian~\cite{Buonanno:1998gg,Buonanno:2000ef,Damour:2001tu,Damour:2008qf,Barausse:2009xi,Khalil:2020mmr}.
Recently, there has been great progress in obtaining these observables directly from the scattering amplitude~\cite{Kosower:2018adc,Maybee:2019jus,Damour:2017zjx,Bjerrum-Bohr:2019kec}. Moreover, in the non-spinning case, Refs.~\cite{Kalin:2019rwq,Kalin:2019inp} developed a dictionary between observables for unbound and bound orbits. 

One prominent connection between physical observables and the scattering amplitude is made via the eikonal phase~\cite{Amati:1990xe}.  There are several studies of this connection, especially in the non-spinning case~\cite{Amati:1990xe,Melville:2013qca,Luna:2016idw,Akhoury:2013yua,KoemansCollado:2019ggb,Cristofoli:2020uzm,DiVecchia:2019myk,DiVecchia:2019kta,Bern:2020gjj,Parra-Martinez:2020dzs}. Refs.~\cite{Guevara:2018wpp,Vines:2018gqi,Siemonsen:2019dsu} verified the applicability of this approach for spinning particles in the special configuration where the spins of the particles are orthogonal to the scattering plane. More recently, Ref.~\cite{Bern:2020buy} conjectured a formula that expresses physical observables in terms of derivatives of the eikonal phase for arbitrary orientation of the spin vectors.  

In this section we extend the analysis of Refs.~\cite{Guevara:2018wpp,Vines:2018gqi,Siemonsen:2019dsu} and~\cite{Bern:2020buy}. Specifically, we start by obtaining the eikonal phase via a fourier transform of our amplitudes.  By restricting to the aligned-spin configuration we obtain a scattering angle which matches that of Ref.~\cite{Guevara:2018wpp,Vines:2018gqi,Siemonsen:2019dsu} when we specialize to the black-hole case. Then, we verify the conjecture of Ref.~\cite{Bern:2020buy} by solving Hamilton's equations for the impulse and spin kick, and relating them to derivatives of the eikonal phase.

The eikonal phase $\chi = \chi_1   
+ \chi_2 + \mathcal{O}(G^3)$ is given by
\begin{align}
\chi_1 &= \frac{1}{4m_1m_2\sqrt{\sigma^2-1}}
\int \frac{d^{2}\bm{q}}{(2\pi)^{2} }e^{-i\bm{q}\cdot\bm{b}}\mathcal{M}^{\rm tree}(\bm{q}) \, , \nn \\
\chi_2 &= \frac{1}{4m_1m_2\sqrt{\sigma^2-1}}
\int \frac{d^{2}\bm{q}}{(2\pi)^{2} }e^{-i\bm{q}\cdot\bm{b}}\mathcal{M}^{\bigtriangleup+\bigtriangledown}(\bm{q}) \, .
\end{align}
Using our amplitudes expressed in the center-of-mass frame (see Eq. (\Ref{fullTheoryM})) we find
\begin{align}
\chi_{1} &= \frac{\xi E G}{|\bm p|} \biggl[
-a^{(0)}_1 \ln \bm b^2 - \frac{2a^{(1,1)}_1}{\bm b^2} (\bm p \times \bm S_1) \cdot \bm b
+a^{(2,4)}_1\left(\frac{2}{\bm b^2} \bm S_{1\perp}^2
-4\frac{(\bm S_{1\perp}\cdot \bm b)^2}{\bm b^4} \right)
\biggr] \,, \nonumber  \\[7pt]
\chi_{2} &= \frac{\pi  \xi E G^2}{|\bm p|} \bigg[
\,\frac{a^{(0)}_2}{|\bm b|}
- \frac{a^{(1,1)}_2}{|\bm b|^3} (\bm p \times \bm S_1) \cdot \bm b
+a^{(2,4)}_2\left(\frac{1}{|\bm b|^3} \bm S_{1\perp}^2
-3\frac{(\bm S_{1\perp}\cdot \bm b)^2 }{|\bm b|^5} \right)\label{Eikonal} \\*
&\hskip 8.0 cm 
\null 
-\left(a^{(2,5)}_2 \bm S_1^2 + a^{(2,6)}_2 (\bm p\cdot \bm S_1)^2
\right)\,\frac{1}{|\bm b|^3} 
\bigg] \,, \nn
\end{align}
where we define $\bm S_{\perp 1} \equiv \bm S_{1} - 
\frac{\bm S_{1} \cdot \bm p} {\bm p ^2} \, \bm p$. 

We may now use the eikonal phase to obtain certain classical observables. We start by considering the aligned-spin kinematics of Ref.~\cite{Guevara:2018wpp,Vines:2018gqi,Siemonsen:2019dsu}. Specifically, we take the spin to be parallel to the orbital angular momentum, and hence orthogonal to the scattering plane. This implies the relations
\begin{equation}
\vS1\cdot \vb=\vS1 \cdot \vp=0.
\end{equation} 
Since the scattering process is confined to a plane,  it can be described by one scattering angle $\theta = \theta_1   
+ \theta_2 + \mathcal{O}(G^3)$, which we obtain as a derivative of the eikonal phase~\cite{Amati:1990xe}
\begin{align}
\theta_i=-\frac{E}{m_1 m_2 \sqrt{\sigma^2-1}}\pd_b \chi_i\,,
\quad i=1,\,2\,,
\end{align}
where $b=|\vb|$.  The novel piece of the 2PM  angle we obtain is quadratic in spin and given by 
\begin{align}
\nonumber \theta_{2,\, \vS1^2}=
%%%%% begin : angle
\frac{3E\pi G^2\vS1^2}{32m_1^2b^4(\sigma^2-1)^2}\Big\{ 
&m_2\left(
6(5\sigma^4-6\sigma^2+1)
+2\C2 (45\sigma^4-42\sigma^2+5)
\right)\\
+&m_1\left(
(65\sigma^4-66\sigma^2+1)
+\C2 (125\sigma^4-138\sigma^2+29)
\right)
\Big\}
%%%%% end : angle
\,.
\end{align}
By specializing to the black-hole case ($\C2=1$) we reproduce the result of Ref.~\cite{Guevara:2018wpp}.

Ref.~\cite{Bern:2020buy} conjectured a formula that directly relates observables in a scattering event with arbitrary spin orientations to the eikonal phase. The observables in question are the impulse $\Delta \bm p$ and spin kick $\Delta \bm S_1$, where
\begin{alignat}{3}
\bm p(t=\infty) &= \bm p + \Delta \bm p \,, \hspace{.98cm} 
\bm p&&(t=-\infty) = \bm p \,, \nn \\
\bm S_1(t=\infty) &= \bm S_1 + \Delta \bm S_1 \,, \quad 
\bm S_1&&(t=-\infty) = \bm S_1 \,.
\end{alignat}
Specifically, by obtaining the impulse and spin kick through $\mathcal{O}(G^2)$ using Hamilton's equations, we find that they may be written as
\begin{align}
\Delta \bm p_\perp   &= -\{ \bm p_\perp, \chi \}
-\frac{1}{2}\,\{\chi, \{\bm p_\perp, \chi\} \}
-\mathcal{D}_{SL}\left(\chi, \{\bm p_\perp, \chi\} \right)
+\frac{1}{2}\,\{\bm p_\perp,\mathcal{D}_{SL}\left( \chi,  \chi \right) \} \,, \nn \\
\Delta \bm S_1 &= -\{ \bm S_1, \chi \}
-\frac{1}{2}\,\{\chi, \{\bm S_1, \chi\} \}
-\mathcal{D}_{SL}\left(\chi, \{\bm S_1, \chi\} \right)
+\frac{1}{2}\,\{\bm S_1,\mathcal{D}_{SL}\left( \chi,  \chi \right) \} \,.
\label{eq:eikonalBrackets}
\end{align}
In Eq.~(\Ref{eq:eikonalBrackets}) we use the definitions 
\begin{align}
\{\bm p_\perp, f \} \equiv -\frac{\partial f}{\partial \bm b}
\,, \hskip 1 cm 
\{\bm S_1,f \}  \equiv \, \frac{\partial f}{\partial \bm S_1} \times \bm S_1 
\,, \hskip 1 cm 
\mathcal{D}_{SL}\left(f, g \right) \equiv 
 - \, \bm S_1\, \cdot \left(\frac{\partial f}{\partial \bm S_1}\, \times\frac{\partial g}{\partial \bm L_b} \right) \, ,
\label{Brackets}
\end{align}
where $\bm L_{\bm b} \equiv \bm b \times \bm p$.
In the above we decompose the impulse as 
\begin{equation}
\Delta \bm p = \Delta p_{\parallel} \frac{\bm p}{|\bm p|} + \Delta \bm p_\perp\,.
\end{equation}
 Eq.~(\Ref{eq:eikonalBrackets}) does not give $\Delta p_{ \parallel}$. Instead, we obtain $\Delta p_{\parallel}$ from the on-shell condition $( \bm p + \Delta \bm p)^2 = \bm p^2$. 
 
Our calculation establishes the conjecture of Ref.~\cite{Bern:2020buy} at the quadratic-in-spin level. 
The fact that the relation holds without modification when we include these higher-in-spin terms is strong indication for its validity in general.
Our calculation further serves as evidence in favor of the surprisingly compact all-order formula that relates the scattering observables to the eikonal phase,
\begin{equation}
\Delta \mathcal{O} = i e^{-i \chi \mathcal{D} } \{ \mathcal{O}, e^{i \chi \mathcal{D} } \}\,,
\label{eq:eikonalAllOrders}
\end{equation}
where for our case $\mathcal{O} = \bm p_\perp$ or $\bm S_1$, and 
$\chi \mathcal{D} g \equiv \chi g + i \mathcal{D}_{SL}(\chi,g)$.

\section{Conclusions}
\label{sec:conclusions}
In this paper we obtained the 2PM-order Hamiltonian that describes the conservative dynamics of two spinning compact objects in General Relativity up to interactions quadratic in the spin of one of the objects.
We followed the approach of Refs.~\cite{Bern:2020buy,Cheung:2018wkq} which was based on scattering amplitudes and EFT.
Along with the results of \cite{Bern:2020buy} for the bilinear-in-the-spins interactions,  this completes the $\mathcal{O}(G^2)$ analysis of quadratic-in-spin effects not including tidal effects.

To construct the Hamiltonian we followed a matching procedure. Ref.~\cite{Cheung:2018wkq} developed this procedure for non-spinning particles, while Ref.~\cite{Bern:2020buy} extended it to the spinning case.  Specifically, we calculated and matched two amplitudes, one in our full theory and one in an EFT.  Ref.~\cite{Bern:2020buy} introduced the full theory to describe the minimal and non-minimal coupling of particles of arbitrary spin to gravity. The Lagrangian contains operators that are in one-to-one correspondence with those of the worldline EFT of \cite{Levi:2015msa}.  The EFT we used captures the dynamics of non-relativistic spinning particles interacting via a potential with unfixed coefficients.  This EFT extended the one of~\cite{Bern:2020buy} to include operators quadratic in the spin of one of the particles.  By matching the amplitudes computed in these two theories,  we fixed these coefficients and hence determined the desired Hamiltonian.

In our calculation we considered effects up to quadratic in the spin of one of the particles, while we took the other particle to be non-spinning. In terms of our full theory, we included the first non-minimal-coupling operator along with the corresponding arbitrary Wilson coefficient $\C2$.  
Unlike the linear-in-spin results, the effects of this operator are not universal and generic bodies are described by different values of $\C2$. As a specific example, $\C2=1$ describes the Kerr black hole.  
For arbitrary values of $\C2$,  we found that the amplitude depends on $q^2 S_1^2$ and $\left(q \cdot S_1 \right)^2$ independently, rather than on the linear combination $q^2 S_1^2 - \left(q \cdot S_1 \right)^2$.  The latter was expected based on an observation in Refs. \cite{Damgaard:2019lfh,Bern:2020buy}.  Recently, Ref.~\cite{Aoude:2020ygw} also remarked that finite-size effects spoil the above expectation. Interestingly, for the Kerr black-hole case ($\C2=1$) the amplitude indeed depends on the linear combination $q^2 S_1^2 - \left(q \cdot S_1 \right)^2$.

The produced conservative Hamiltonian enables the calculation of observables pertaining to binary systems of spinning black holes or neutron stars.  For example, one may study bound states of the binary by choosing suitable initial conditions. In this paper we chose to compute scattering observables instead, which may be used in the construction of important phenomenological models as the effective one-body Hamiltonian~\cite{Buonanno:1998gg,Buonanno:2000ef,Damour:2001tu,Damour:2008qf,Barausse:2009xi,Khalil:2020mmr}.  Specifically,  by solving Hamilton's equations we obtained the relevant impulse and spin kick. In this way we verified the conjecture of Ref. \cite{Bern:2020buy}, which expresses these observables in terms of the eikonal phase via the simple compact formula in  Eq.~(\ref{eq:eikonalAllOrders}). The existence of such a formula has intriguing implications in classical mechanics.  Specifically, it hints towards a formalism that bypasses using Hamilton's equations, and directly expresses the observables in terms of derivatives of a single function of the kinematics.

In order to establish the validity of our result for the quadratic-in-spin two-body Hamiltonian,  we performed several checks against the literature.  We did this by comparing at the level of the gauge-invariant amplitudes in the regime where they overlap.  Firstly,  we verified that our amplitude expanded in velocity matches the one calculated using the Hamiltonian of Ref.~\cite{Levi:2016ofk}, which was obtained in the PN approximation. Secondly,  by expanding our amplitude in the test-body limit we found agreement with the amplitude obtained by the Hamiltonian of Ref.~\cite{Vines:2016unv}. As a third check, we computed the scattering angle for the kinematic configuration where the spin vector is aligned with the orbital angular momentum of the system and confirmed that it reproduces the one of Ref. \cite{Guevara:2018wpp} for the BH case, $\C2=1$. Finally, we compared the impulse in Eq. (\Ref{eq:eikonalBrackets}) with the one given in Ref. \cite{PortoPMspin} in covariant form and found agreement. 

Our calculation serves as evidence that the formalism of Ref.~\cite{Bern:2020buy} can capture the effects of non-minimal coupling to gravity.  Therefore, an obvious future direction is to extend this analysis to include more powers of spin. Moreover, a number of pressing questions remain interesting and unanswered. These include the extension of these methods to higher PM orders, the proof of the relation between classical scattering observables and the eikonal phase, along with potential extensions of this relation to bound-orbit observables.

\subsection*{ Acknowledgments:}
We thank Zvi Bern,  Radu Roiban,  Chia-Hsien Shen and Fei Teng for collaboration in related topics, and Donal O'Connell for discussions. We thank Jan Steinhoff and Justin Vines for several helpful discussions, and for supplying ancillary files for the post-Newtonian Hamiltonian of Ref. \cite{Levi:2015msa}, as well as a simpler form of the test-mass Hamiltonian based on Ref.  \cite{Vines:2016unv}. We also thank the authors of \cite{PortoPMspin} for sharing their results for the linear impulse in advance of its publication.
DK and AL are supported by the U.S. Department of Energy (DOE) under award number DE-SC0009937, and by the Mani L. Bhaumik Institute for Theoretical Physics.  

%\bibliography{ref}{}
%\bibliographystyle{utcaps}

%% Regarding bibtex
% One way to make arxiv refs work and to have it sort them automatically is to use utcaps.bst
% You can find it here:
% https://arxiv.org/help/hypertex/bibstyles

\providecommand{\href}[2]{#2}\begingroup\raggedright\endgroup

\end{document}